\documentclass{article}

\usepackage[english]{babel}
\usepackage{authblk}
\usepackage{longtable}

\usepackage{graphicx}%
\usepackage{hyperref}

\usepackage{multirow}%
\usepackage{amsmath,amssymb,amsfonts}%
\usepackage{amsthm}%
\usepackage{mathrsfs}%
\usepackage[title]{appendix}%
\usepackage{xcolor}%
\usepackage{textcomp}%
\usepackage{manyfoot}%
\usepackage{booktabs}%
\usepackage{algorithm}%
\usepackage{algorithmicx}%
\usepackage{algpseudocode}%
\usepackage{listings}%

\usepackage[utf8]{inputenc}
\usepackage{longtable}

\usepackage{parskip}
\usepackage{graphicx}
\usepackage{setspace}
\usepackage{xcolor}
\usepackage[table]{xcolor} % Per colorare la tabella
\usepackage{booktabs} % Migliore formattazione

\usepackage{amsmath}
\usepackage{graphicx}
\title{Socioeconomic Determinants of the COVID-19 Infodemics}

\author[1, 2, 3]{Anna Bertani}
\author[1]{Alessandro Cortese}
\author[4]{Federico Pilati}
\author[5]{Pierluigi Sacco}
\author[1]{Riccardo Gallotti}

\affil[1]{Fondazione Bruno Kessler, Via Sommarive, 18, 38123, Povo TN}
\affil[2]{University of Trento, Department of Information Engineering and Computer Science, Via Sommarive 9, 38123, Povo TN}

\affil[4]{Department of Social and Political Sciences, University of Bologna, Strada Maggiore 45, Bologna, 40125, Italy}

\affil[5]{Department of Neurosciences, Imaging and Clinical Sciences, University of Chieti-Pescara, Via Luigi Polacchi, 11, Chieti, 66100, Italy}
\affil[3]{Centre for Sociology of Humans and Machines (SOHAM), Trinity College Dublin, 3-5 Foster Place, D02 YT92, Dublin, Ireland}

\begin{document}
\maketitle

\begin{abstract}
The COVID-19 pandemic has been accompanied by an ``infodemic'' of misinformation that impedes effective public health responses. This study examines relationships between socioeconomic factors and infodemic risk patterns across 37 OECD countries using Twitter data from 2020-2022. Employing dimensionality reduction techniques on 20 socioeconomic indicators, we identify complex correlations with infodemic measures that evolve throughout the pandemic. Countries exhibit distinct clustering in their infodemic profiles that transcend conventional socioeconomic categorizations. We find that dynamic information behaviors dominate initial crisis responses, while stable socioeconomic conditions become more influential as the pandemic progresses. News media diet diversity emerges as a significant protective factor, with pluralistic information ecosystems demonstrating greater resilience against misinformation. Additionally, institutional stability correlates strongly with reduced infodemic volatility over time. These findings highlight how infodemics are embedded within broader socioeconomic contexts, providing foundations for targeted interventions to build societal resilience against misinformation during future health emergencies.
\end{abstract}

\section{Introduction}

The COVID-19 pandemic has presented unprecedented challenges to global public health systems, economies, and social structures. Amid this crisis, another parallel phenomenon has emerged and gained increasing recognition: what the World Health Organization has termed an ``infodemic'' -- an overabundance of information, both accurate and inaccurate, that makes it difficult for people to find trustworthy sources and reliable guidance when needed \cite{zarocostas2020fight} \cite{world2020coronavirus}. This concept of information disorder during public health emergencies is not entirely new; historical instances such as the HIV/AIDS epidemic \cite{kalichman2009denying}, H1N1 outbreak \cite{tausczik2012public}, and Ebola crisis \cite{allgaier2015communication} have all been accompanied by waves of misinformation. However, the scale, speed, and impact of the COVID-19 infodemic has been dramatically amplified by the digital transformation of our information ecosystem, particularly the rise of social media platforms as primary news sources for significant portions of the global population \cite{nielsen2020navigating}.

The consequences of the infodemic extend far beyond mere confusion. Studies have linked exposure to COVID-19 misinformation with reduced compliance with public health measures \cite{bridgman2020causes}, increased vaccine hesitancy \cite{loomba2021measuring}, exacerbated social tensions \cite{romer2020conspiracy}, and even physical harm from dangerous remedies promoted through false claims \cite{islam2020covid}. As argued by the author of \cite{gallotti2020assessing}, infodemics can substantially increase epidemic spread by undermining confidence in scientific evidence and public health authorities, thereby hampering coordinated social responses needed for effective pandemic control.

While early research on misinformation often emphasized technological factors such as algorithmic amplification \cite{burbach2019bubble} \cite{tornberg2018echo} or the absence of traditional gatekeepers in digital media \cite{bode2018see}, a growing body of literature recognizes infodemics as complex socio-technical phenomena embedded within broader social, political, and economic structures \cite{starbird2019disinformation} \cite{bennett2018disinformation}. This shift in perspective is crucial, as it suggests that societal vulnerability to information disorders cannot be addressed through technological solutions alone, but requires an understanding of the underlying socioeconomic conditions that create fertile ground for misinformation to flourish.

Recent studies have begun to explore how various social factors might influence misinformation dynamics. \cite{humprecht2020resilience} found that countries with strong public service media, higher levels of trust in news, and lower levels of political polarization showed greater resilience against online misinformation. \cite{bavel2020using} highlighted how existing social divisions and identities shape individuals' receptivity to misinformation. \cite{vosoughi2018spread} demonstrated that false news travels faster and reaches more people than accurate information, particularly when it evokes emotional responses of fear, disgust, or surprise. More recent work has emphasized the importance of comparing cascades of equal size before drawing con- clusions on content-specific differences, in order to disentangle structural effects from topical ones \cite{juul2021comparing}. These studies collectively suggest that social, cultural, and economic factors play critical roles in determining how information and misinformation spread through populations.

However, empirical research systematically examining the relationship between comprehensive sets of socioeconomic indicators and infodemic vulnerability across multiple countries remains scarce. Much of the existing literature focuses on single countries, such as Brazil \cite{harb2022analysis}, or on individual-level factors rather than structural conditions \cite{pennycook2019lazy} \cite{bronstein2019belief}. While these studies provide valuable insights, they offer only a partial understanding of how broader societal characteristics might influence a nation's overall information environment during a global crisis.

Social media platforms have become central to crisis communication during the COVID-19 pandemic \cite{cinelli2020covid} \cite{singh2020first}, serving as critical channels for rapid information dissemination from health authorities \cite{chan2020social}, spaces for public discourse and sense-making \cite{wicke2020framing}, and unfortunately, vectors for the spreading of misinformation \cite{brennen2020types}. Twitter, in particular, has emerged as a significant platform for COVID-19 communication due to its real-time nature, public accessibility, and adoption by key stakeholders including government officials, health organizations, journalists, and public health experts \cite{chen2020tracking} \cite{sacco2021emergence}.

Analyses of Twitter data have yielded important insights into information flow patterns during the pandemic. Various studies have examined the prevalence of misinformation in COVID-19 related tweets \cite{forati2021geospatial} \cite{unlu2025tracing}, the role of bots in spreading false claims \cite{suarez2025unraveling} \cite{unlu2024unveiling}, emotional responses to pandemic developments \cite{storey2024text} \cite{shen2024online}, and the evolution of topics in public discourse \cite{batool2024enhanced} \cite{alshanik2025unveiling}. \cite{gallotti2020assessing} developed a quantitative measure of ``infodemic risk'' based on the reliability of news sources shared on Twitter, creating a framework for cross-national comparisons of information quality during the pandemic.

While these studies have advanced our understanding of digital communication dynamics during the COVID-19 pandemic, they have generally not systematically explored how these patterns correlate with the underlying socioeconomic characteristics of different countries. This represents a significant gap in our knowledge, as understanding these relationships could provide crucial insights for developing more effective strategies to combat infodemics.

The broader literature on media systems and information environments suggests numerous ways in which socioeconomic factors might influence a society's vulnerability to infodemics. Educational attainment, particularly levels of scientific and digital literacy, may affect a population's abilities to critically evaluate health information \cite{aydinlar2024awareness} \cite{heiss2024debunking}. Political polarization can lead to motivated reasoning and selective exposure to ideologically congruent information, regardless of its accuracy \cite{van2018partisan} \cite{van2024political}. Economic inequality may exacerbate information disparities and create segmented information environments \cite{rasanen2021online} \cite{mostagir2023social}. Media freedom and pluralism shape the diversity and quality of available information sources \cite{humprecht2020resilience}.

Countries differ substantially across these dimensions, suggesting potential variation in their vulnerability to infodemics. However, these factors likely operate in complex, interrelated ways rather than in isolation. For instance, Hallin and Mancini's \cite{hallin2004comparing} influential work on comparative media systems demonstrated how media institutions are deeply embedded in broader political and social structures, creating distinctive patterns across different types of democracies. Similarly, information ecosystems during crises likely reflect complex interactions between multiple socioeconomic dimensions rather than simple correlations with individual factors.

Despite the growing recognition of infodemics as a socially embedded phenomenon, several important research gaps remain. First, few studies have systematically examined how comprehensive sets of socioeconomic indicators relate to infodemic patterns across multiple countries. Second, existing research has largely overlooked the temporal dynamics of infodemics, i.e., how they evolve over time and whether their relationship with socioeconomic factors changes during different phases of a crisis. Third, the role of media diet diversity as a potential mediating factor between socioeconomic conditions and infodemic vulnerability remains relatively underexplored.

Our study addresses these gaps by investigating the relationship between 20 socioeconomic indicators and patterns of infodemic risk across 37 OECD countries (with the exception of Luxembourg due to missing indices) during the COVID-19 pandemic. Using dimensionality reduction techniques, we identify underlying structures in our socioeconomic data and examine how these dimensions correlate with various measures of infodemic risk at different time points. Additionally, we explore how the diversity of news media diets, measured using Shannon entropy \cite{bertani2024decoding}, relates to both socioeconomic factors and infodemic vulnerability.

Specifically, our research aims to answer the following questions. First, what underlying dimensions can be identified in the socioeconomic indicators of OECD countries, and how do countries cluster along these dimensions? Moreover, do countries exhibit distinct clusters based on their temporal patterns of infodemic risk during the COVID-19 pandemic? In addition, how do socioeconomic factors correlate with different measures of infodemic risk (overall risk, dynamic risk, and unreliability), and do these correlations change over time? Furthermore, what role does news media diet diversity play in mediating the relationship between socioeconomic factors and infodemic vulnerability? And finally, how does infodemic volatility relate to socioeconomic factors, and does this relationship evolve over the course of the pandemic?

By addressing these questions, our study contributes to both theoretical understanding and practical approaches to the management of infodemics. Theoretically, we advance a more nuanced framework for understanding infodemics as phenomena embedded within complex socioeconomic contexts, moving beyond purely technological or individual-level explanations. Practically, the identification of socioeconomic correlates of infodemic vulnerability can inform more targeted and effective interventions to build societal resilience against misinformation during future public health emergencies.

In the following sections, we describe our methodology for collecting and analyzing Twitter data to measure infodemic risk across countries, as well as our approach to gathering and processing socioeconomic indicators. Subsequently, we present results from our dimensional analysis of socioeconomic factors, our identification of infodemic risk clusters, our examination of correlations between socioeconomic dimensions and infodemic measures, and our exploration of the role of news media diet diversity. Finally, we discuss the implications of our findings for understanding and addressing infodemics in the context of global public health crises.

\section{Methods}\label{sec11}
\subsection{Twitter dataset}

Our analysis relies on data collected through the COVID-19 Infodemic Platform  \cite{covid19infodemics}, a comprehensive repository of Twitter (now $\mathbb{X}$) messages posted throughout the COVID-19 pandemic \cite{gallotti2020assessing}. This dataset encompasses messages containing URLs, each evaluated for reliability through comparison with various publicly accessible databases covering scientific and journalistic sources. We specifically employed reliability classifications from MediaBiasFactCheck \cite{mediabiasfact}, an organization that maintains an extensive and regularly updated database evaluating the ideological leanings and factual accuracy of media outlets through both quantitative metrics and qualitative assessments. 

The URLs in our dataset were classified into seven distinct categories: reliable sources (Mainstream Media, Scientific Journals), unreliable sources (Satire, Clickbait, Political, Fake/Hoax, Conspiracy/Junk Science), and unclassified sources. Building on this classification system, we employed three metrics developed by \cite{gallotti2020assessing} to quantify different dimensions of infodemic risk. The Infodemic Risk Index (IRI) quantifies the rate at which a user is exposed to unreliable news.  To estimate the exposure to unreliable news ($E_u$), we aggregate the number of followers who are potentially reached by tweets containing unreliable news. Conversely, the exposure to reliable news ($E_r$) is given by aggregating the number of followers potentially reached by tweets containing reliable sources. The IRI is computed as \[IRI = \frac{E_U}{E_U + E_R}\] 
The \emph{Dynamic Infodemic Risk Index} is the likelihood that a user engages with online messages pointing to unreliable sources. The index is calculated as above, but $E_U$ and $E_R$ represent the total number of interactions (retweets, replies, quotes) of messages containing, respectively, unreliable and reliable news. Finally, the \emph{News Unreliability Index} is the proportion of unreliable news over the total number of tweets containing links pointing to any information sources. 

\subsection{Socioeconomic indicators dataset}

To examine the relationship between infodemic risk and socioeconomic factors, we compiled an extensive dataset of socioeconomic indicators for 37 OECD countries \cite{oecdMembers}. This selection ensured data consistency and reliability across all indicators. Our initial compilation encompassed 14 socioeconomic indicators with complete data for all countries in our sample. We subsequently enriched this dataset with six additional indicators from the OECD's comprehensive database, resulting in a final set of 20 indicators spanning multiple dimensions of social, political, and economic development.

The selected indicators capture diverse aspects of national development: democratic quality (EIU Democracy Index, EIU Political Participation); civil liberties (Freedom House Freedom Index); societal stability (Global Peace Index, Global Terrorism Index); economic prosperity (GDP per capita); educational attainment (percentage of adults with tertiary education, percentage of foreign students, PISA mathematics and science scores); social welfare (social security expenditure, social spending); institutional trust (trust in government); media freedom (Press Freedom Index); digital participation (E-Government Participation Index); demographic structure (median age); social cohesion (political polarization, social polarization); digital development (internet use percentage); and economic inequality (Gini coefficient). A detailed description of each indicator, including its source and measurement methodology, is provided in Appendix A.

\subsection{Dimensionality Reduction Techniques}
The \emph{curse of dimensionality} arises in high-dimensional datasets where the number of features greatly exceeds the observations, causing issues such as sparsity, computational inefficiency, overfitting, and challenges in visualization. Dimensionality reduction techniques mitigate these problems by mapping data to a lower-dimensional space while preserving essential information. This enables an easier visualization of patterns and clusters in two or three dimensions and reduces noise and redundancy in features. By simplifying the dataset, these methods enhance interpretability and improve the performance of analytical models on complex data.

\subsubsection{UMAP}
We employed Uniform Manifold Approximation and Projection (UMAP) as our primary dimensionality reduction technique. UMAP represents a state-of-the-art algorithm that effectively visualizes high-dimensional data while maintaining an appropriate balance between local and global structure preservation  \cite{mcinnes2018umap}. UMAP is moreover computationally fast and scales well with regard to both dataset size and dimensionality. It works by approximating the underlying manifold of the data and projecting it into a low-dimensional space, typically two or three dimensions, through two key phases: graph construction and graph projection. In the first phase, a high-dimensional graph is created using a fuzzy simplicial complex, where the edge weights indicate the connection strength between the data points. The second phase optimizes this graph to a low-dimensional Euclidean space using cross-entropy. UMAP’s tunable hyperparameters, such as n\_neighbors, min\_dist, and the distance metric, enable users to control the embedding’s granularity and structure, though careful tuning is required to avoid distortions. Compared to other techniques, UMAP excels in capturing non-linear patterns and is faster than t-SNE, while also supporting a variety of distance metrics \cite{becht2018evaluation}. 

\subsection{Principal Component Analysis}

Principal Component Analysis (PCA) is a statistical technique used to reduce the dimensionality of a dataset while preserving as much of its variability as possible \cite{jolliffe2016principal}. It does so by transforming the original, possibly correlated variables into a smaller set of uncorrelated variables called principal components, which are ordered by the amount of variance they explain. By projecting the data onto these components, PCA simplifies complex datasets, making patterns easier to identify and analyze while minimizing information loss.

\subsection{K-means}

K-means clustering is a popular and straightforward method for partitioning data into 
k distinct and non-overlapping clusters. The algorithm works by minimizing the total within-cluster variation, a measure of similarity of the data points within each cluster, often calculated using the squared Euclidean distance. The process begins with an initial random assignment of cluster labels to the data points. Then, it iteratively refines the clusters by computing the centroids of each cluster and reassigning each data point to the cluster with the closest centroid. This optimization process continues until convergence, ensuring that the objective function (minimizing within-cluster variance) decreases at each step.

While k-means is computationally efficient, intuitive, and widely applicable, it has significant limitations. One major drawback is its sensitivity to the initialization of cluster centroids; different initializations can lead to varying outcomes, as the algorithm is only guaranteed to find a local optimum. Another challenge is the need to specify the number of clusters (k) in advance, which is often not straightforward. Choosing an incorrect value for k can lead to poor results, such as clustering noise or failing to capture meaningful groupings in the data.
Several heuristic methods, like the elbow method and silhouette score, have been developed to guide the selection of k. The elbow method involves plotting the within-cluster sum of squares against different values of k and identifying the point where the decrease in variance begins to level off, forming an ``elbow''. The silhouette score evaluates how well each data point fits within its assigned cluster compared to other clusters. However, these methods are not always consistent or reliable, and there is no definitive consensus on the best approach to determine k. Despite these challenges, k-means remains a robust and widely used clustering technique, especially for large datasets where simplicity and speed are critical.

\subsection{News Media Diet}

To better understand the relationship between the Infodemic Risk Index and socioeconomic indicators, we used entropy as a measure to capture the variety of news outlets in each country. We recall here the concept of news media diet, as discussed in \cite{bertani2024decoding}.  With the term \emph{news media diet}, we refer to the collection of web domains or URLs shared in each country.  In particular, we employed the temporal-uncorrelated entropy, computed as the Shannon entropy  \[
S_{unc} = -\sum_{j=1}^{N_i} p(x_j) \cdot \log_2(p(x_j))
\]
where $p(x_{i})$ is the historical probability that a web-domain $x_{i}$ was shared by user \emph{i} and $N(i)$ is the total number of users. This measure enables us to characterize the heterogeneity of the news media diet.

\subsection{Volatility}

To assess the stability of information environments across countries, we calculated the volatility of the Infodemic Risk Index, defined as the standard deviation of the IRI over time for each country. Initial analysis revealed a strong negative correlation between the number of tweets posted daily and the volatility of the index, indicating that countries with higher tweet volumes exhibited more stable IRI values over time. This correlation follows an expected statistical relationship, where the standard deviation scales with the square root of the sample size  $\sigma \approx (N^{0.5})$. To account for this effect and obtain a measure of volatility independent of sample size, we adjusted our volatility metric by multiplying the standard deviation by the square root of the average daily tweet count: $sigma^* = \sigma* \langle N \rangle^{-0.5}$. This adjustment enabled more meaningful cross-national comparisons of infodemic volatility by controlling for differences in tweet volume across countries.

\section{Results}\label{sec2}

This study explores the relationship between infodemic risk indices and 20 socioeconomic indicators across 37 OECD countries, employing dimensionality reduction techniques to identify underlying patterns and correlations. Our analyses reveal complex relationships between the socioeconomic fabric of nations and their vulnerability to infodemics during the COVID-19 pandemic.

Firstly, we used UMAP, a dimensionality reduction method (see Methods), to project data from socioeconomic indices and countries in a two-dimensional space, allowing us to discern the underlying structure of the OECD data more effectively. 

\begin{figure}[!ht]
\centering
\includegraphics[width=1\textwidth]{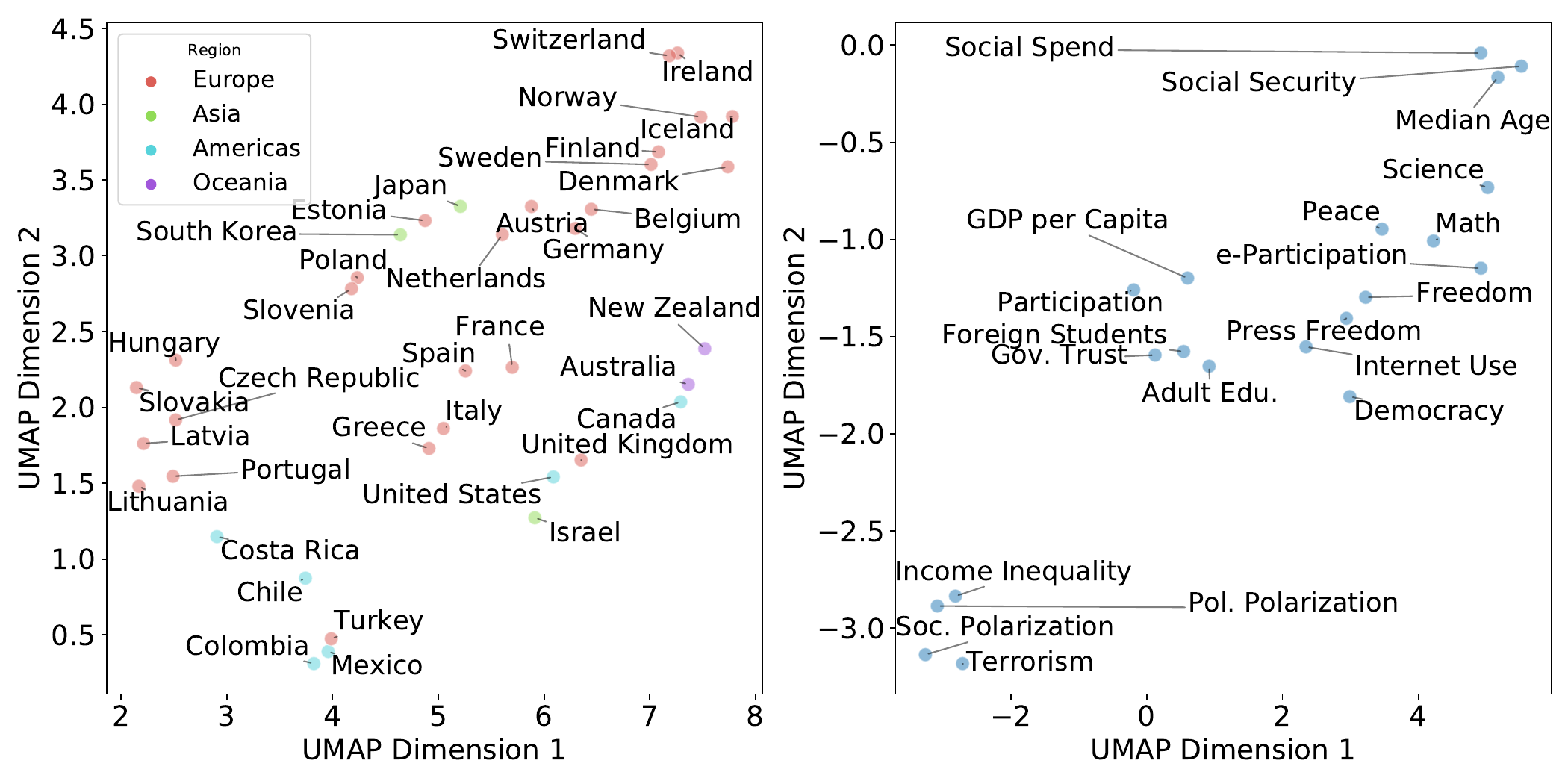}
\caption{{\bf UMAP visualization of OECD countries and the socioeconomic indices} UMAP visualization of the OECD* countries based on the selected indices (left) and of the selected socioeconomic indices (right). The employed hyperparameters are n neighbors = 5 and min dist = 0.01.}\label{fig1}
\end{figure}

The application of UMAP dimensionality reduction to socioeconomic data, presented in Figure \ref{fig1}, reveals distinct structural patterns that transcend simple geographic or economic categorizations. The country projection (Figure \ref{fig1}, left panel) demonstrates clear geopolitical and cultural clustering: Scandinavian and Central European nations congregate in the upper right quadrant, Anglosphere countries position to the right, while South American nations cluster in the bottom left. This visualization captures deeper institutional and cultural similarities that unite countries beyond their geographic proximity, reflecting historical patterns of institutional development and socioeconomic evolution.

The arrangement of socioeconomic indicators (Figure \ref{fig1}, right panel) along a curvilinear distribution provides equally meaningful insights. Welfare and demographic indicators (median age, social security, social spending) position at the upper region, followed by democracy, education, and economic development indicators in the central area. Notably, indicators of social and political polarization appear distinctly separated in the bottom left, alongside measures of income inequality and terrorism. This separation suggests that polarization constitutes a distinct dimension of social development that can vary independently of traditional development metrics. The effectiveness of UMAP in discerning these multidimensional relationships shows how certain societal characteristics cluster together, forming coherent socioeconomic profiles that may influence information environments during crises.

\begin{figure}[!ht]
\centering
\includegraphics[width=1\textwidth]{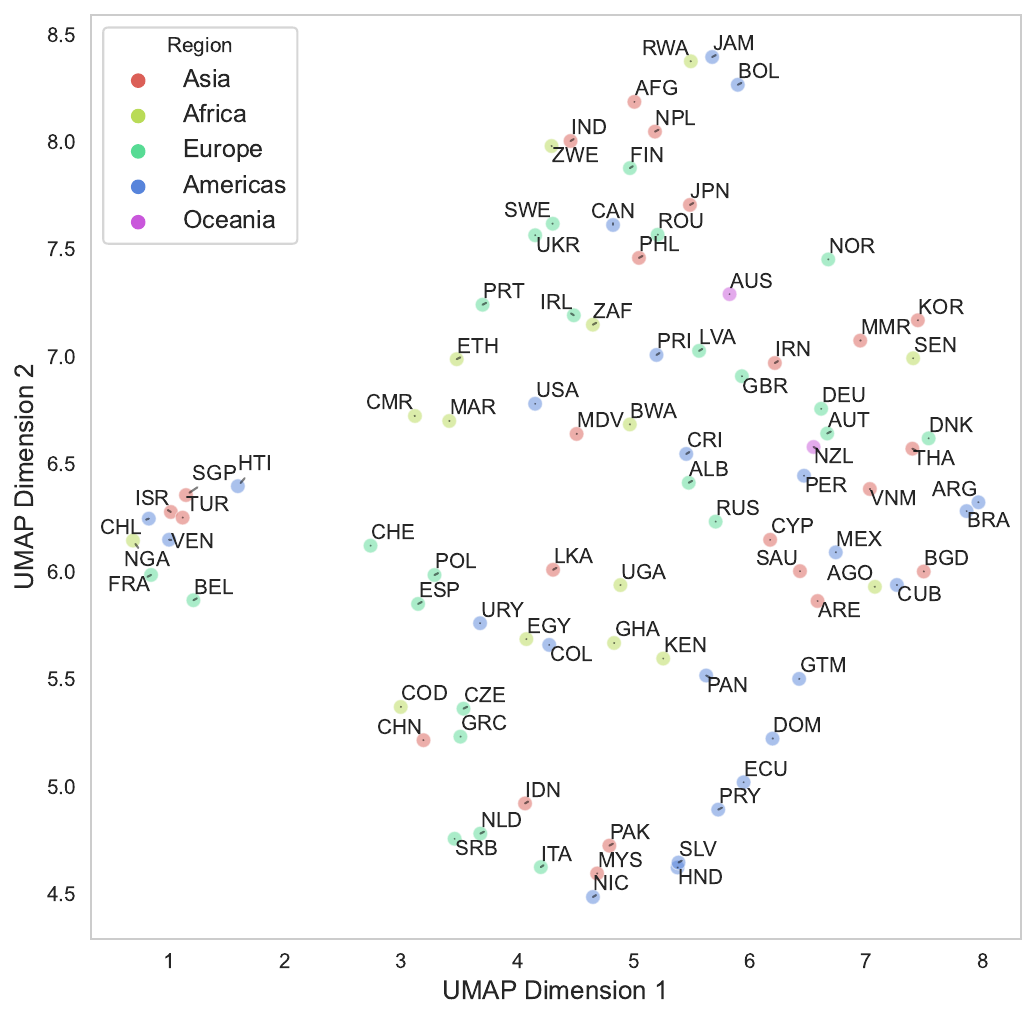}
\caption{UMAP projection of the IRI time series of the countries featuring at least 500’000 tweets in the relevant time frame. The selected parameters are n neighbors = 5 and min dist = 0.1. The 9 countries belonging to Cluster 1 are grouped on the left, whereas the remaining ones belong to Cluster 2.}\label{fig2}
\end{figure}

Our temporal analysis of infodemic risk patterns across countries yielded significant findings regarding how information environments evolved during the pandemic. Figure \ref{fig2} illustrates the UMAP projection of countries based on their Infodemic Risk Index (IRI) time series, revealing two distinct clusters that transcend conventional socioeconomic categorizations. The first cluster, comprising nine countries (Belgium, Chile, France, Haiti, Israel, Nigeria, Singapore, Turkey, and Venezuela), appears isolated on the left side of the projection, while the second, larger cluster encompasses the remaining countries on the right.

\begin{figure}[h]
\centering
\includegraphics[width=1\textwidth]{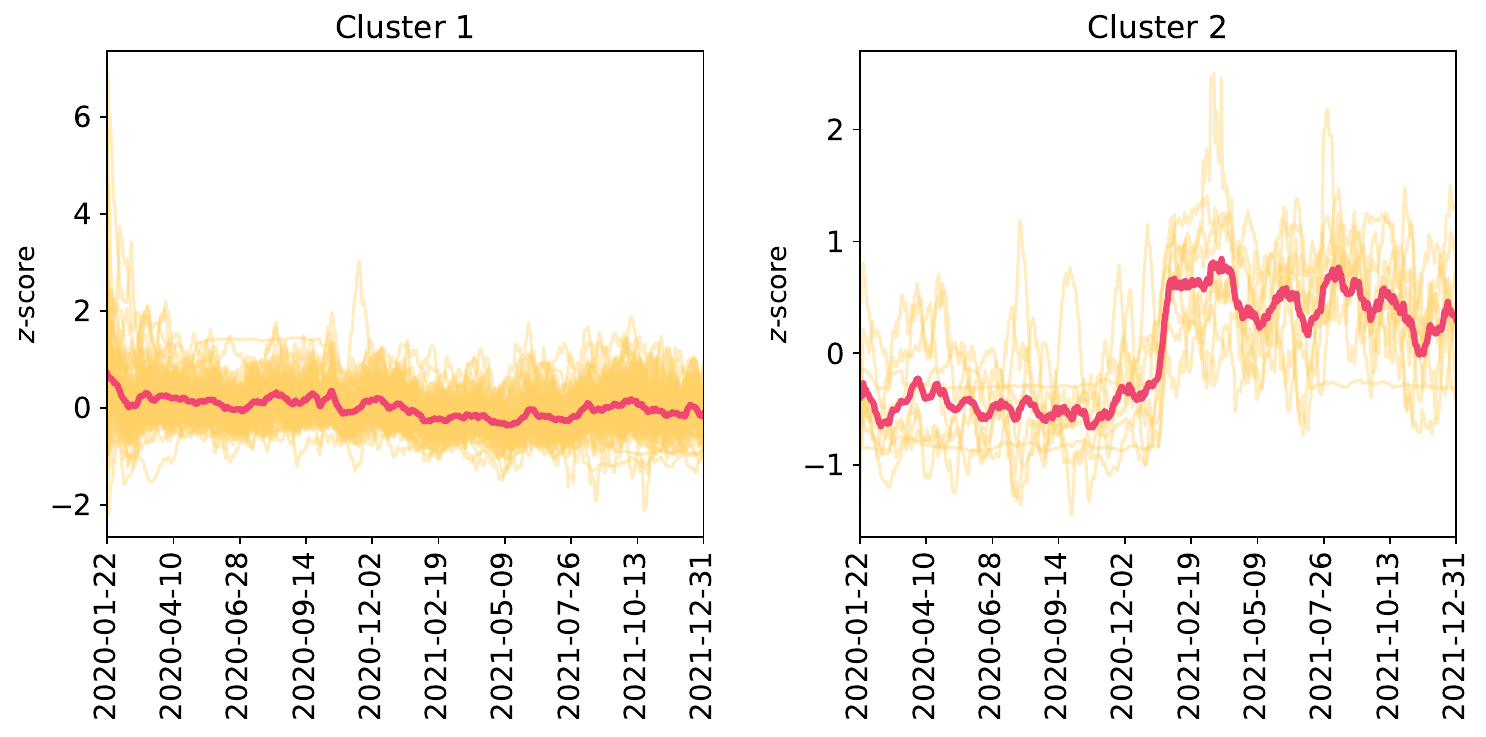}
\caption{k-means clustering with Euclidean distance and UMAP preprocessing on the IRI time series for the 88 selected countries. The UMAP parameters employed were n neighbors = 5, min dist = 0.1 and n components = 102. The cluster average is plotted in red. All the time series are smoothed using a 30-day rolling average.
}\label{fig3}
\end{figure}

 To better understand these distinctive clustering patterns, we applied a \emph{k-means} clustering to the temporal evolution of the Infodemic Risk Index (Figure \ref{fig3}). The analysis confirms two clearly differentiated infodemic risk profiles: Cluster 2 maintains a relatively stable risk profile throughout the observed period, whereas Cluster 1 exhibits a pronounced spike in risk during early 2021, with z-scores fluctuating between -0.5 and 0.5. This temporal divergence suggests that certain combinations of socioeconomic factors may create vulnerability to specific types of infodemic events, potentially related to vaccine rollout controversies that emerged during this period. The heterogeneity of countries within Cluster 1, spanning diverse regions, development levels, and political systems, indicates that infodemic vulnerability follows complex patterns that are not reducible to single socioeconomic dimensions. 

\begin{figure}[!ht]
\centering
\includegraphics[width=1\textwidth]{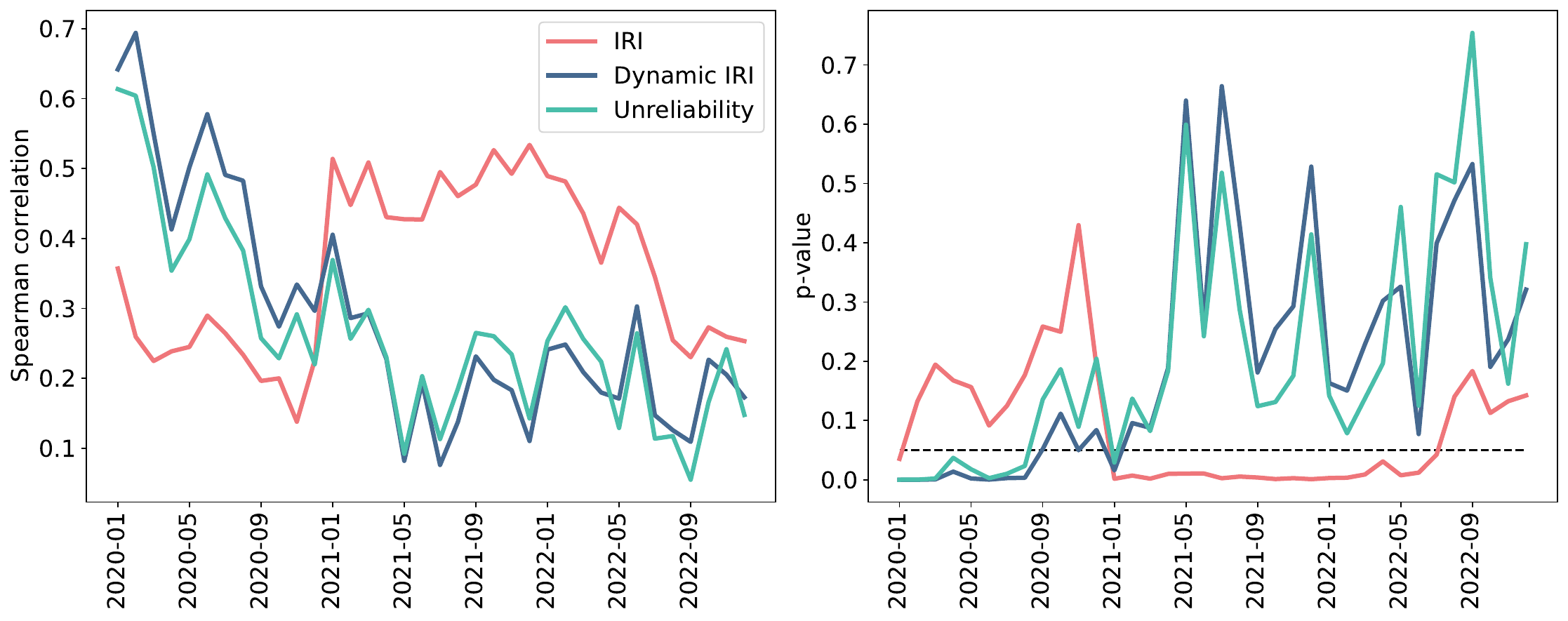}
\caption{The first panel shows the Spearman correlation of the first Principal Component and the monthly average of the infodemic metrics over time for OECD* countries. The second panel displays the p-value of the corresponding correlation with a black dashed line representing the significance level of 5\%.}\label{fig4}
\end{figure}

To investigate how socioeconomic factors relate to infodemic dynamics over time, we extracted the first principal component of our socioeconomic indicators and analyzed its correlation with three infodemic metrics (IRI, Dynamic Risk Index and Unreliability Index) over the course of the pandemic (Figure \ref{fig4}). This principal component represents the linear combination of socioeconomic indicators that captures the maximum variance in our dataset, serving as a composite proxy for socioeconomic development.

The first panel of Figure \ref{fig4} reveals distinct temporal patterns in these correlations. During the initial phase of the pandemic (2020), the Dynamic Risk Index and the Unreliability Index demonstrate stronger correlations with socioeconomic factors, with coefficients reaching approximately 0.7. However, as the pandemic progresses into 2021, a marked shift occurs: these correlations diminish while the correlation between socioeconomic factors and the overall Infodemic Risk Index increases substantially, stabilizing around 0.5 through May 2022.

The correlation of Dynamic Infodemic Risk and Unreliability Index follows the same pattern and decreases with the beginning of 2021. The second panel, depicting p-values for these correlations, confirms the statistical significance of these temporal patterns. Two distinct phases emerge: an early phase characterized by significant correlations between socioeconomic factors and measures of information dynamics (Dynamic Risk) and content reliability (Unreliability Index), followed by a later phase where overall infodemic risk becomes more statistically significant and more strongly associated with socioeconomic conditions. This evolution suggests that different aspects of information environments become salient at different stages of a crisis, with initial responses dominated by dynamic information behaviors that later crystallize into more stable patterns of overall risk.
  
Interestingly, our analysis shows that the public conversation on COVID-19 evolved distinctly over the three-year pandemic period. As illustrated in Figure \ref{fig2}, countries clustered into two clearly differentiated groups based on their infodemic risk profiles. These clusters reflect substantive differences in how information environments responded to the pandemic across nations. Figure \ref{fig3} further demonstrates that these differences manifested in distinct temporal patterns, with one cluster exhibiting marked fluctuations in Infodemic Risk Index values while the other maintained relatively stable information environments throughout the crisis period.

\begin{figure}[!ht]
\centering
\includegraphics[width=1\textwidth]{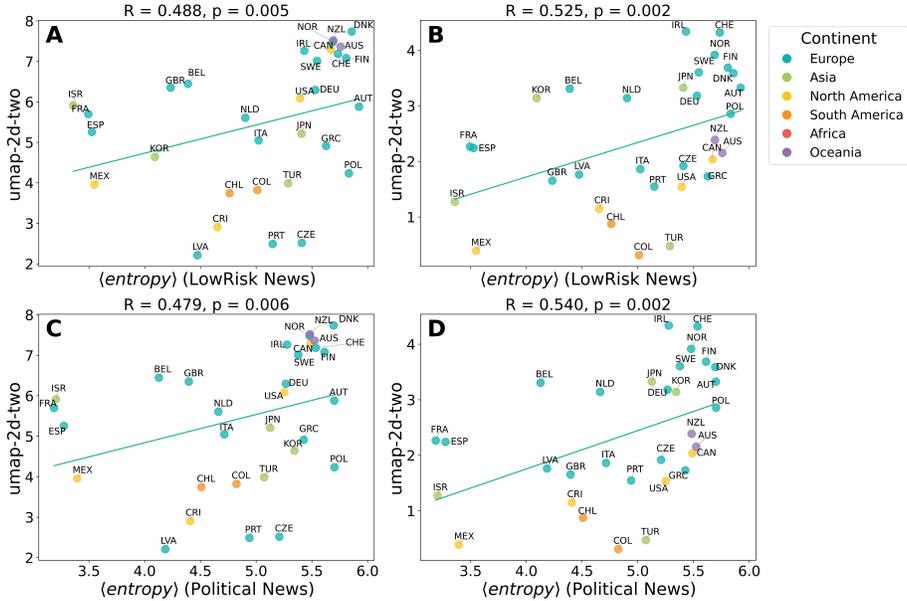}
\caption{Correlation between the news media diet and the UMAP of the socioeconomic indices. Panels A and B show the correlation between the socioeconomic indicators with the entropy of all the domains classified as low-risk (satire, clickbait and political) with a spearman correlation of 0.49 in 2020 and 0.52 in 2021. Panels C and  D represent the correlation between the socioeconomic indices and the news classified as political, with a spearman correlation value respectively of 0.48 and 0.54.}\label{fig5}
\end{figure}

 Based on these results, we decided to add another variable in order to explain the variability of the Infodemic Risk Index with respect to the socioeconomic indicators. Figure \ref{fig5} shows the correlation between the two-dimensional space of the socioeconomic indicators and the diversity of  the news media diet of each country analyzed. 
 
 Figure \ref{fig5} illustrates the correlation between the news source diversity, measured as the Shannon entropy of shared web domains, and the two-dimensional UMAP projection of socioeconomic indicators. Panels A and B report a robust positive correlation between the diversity of low-risk news consumption (encompassing satire, clickbait, and political content) and both dimensions of the socioeconomic space, with Spearman coefficients of 0.48 for the first dimension and 0.52 for the second, both statistically significant.

This relationship strengthens further when focusing specifically on political news diversity (Panels C and D), particularly with respect to the second dimension. These findings suggest that countries positioned in socioeconomic contexts characterized by stronger democratic institutions, higher education levels, and greater economic development tend to exhibit more diverse news consumption patterns. The pronounced correlation with the second dimension indicates that institutional quality and governance factors may be particularly influential in fostering media pluralism. The stronger correlation observed for political news diversity suggests that the pattern of engagement with political information may be especially sensitive to underlying socioeconomic conditions. This reinforces the idea that open, pluralistic societies with well-functioning democratic structures encourage a broader and more varied media landscape, whereas lower entropy may be indicative of more concentrated or polarized information environments.

\begin{figure}[h]
\centering
\includegraphics[width=\textwidth]{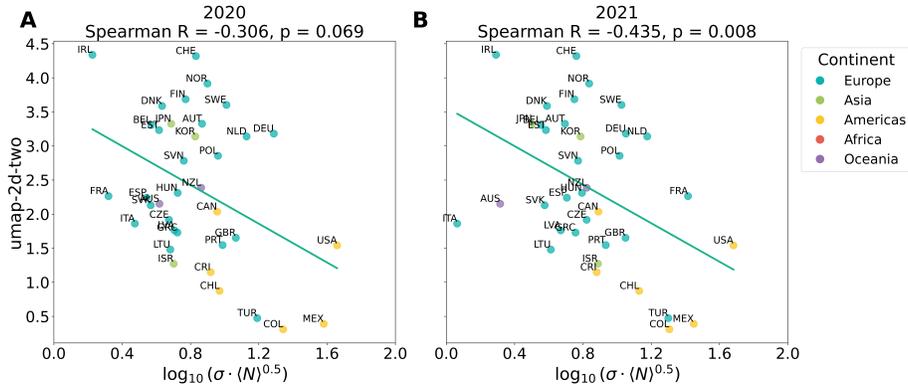}
\caption{Correlation between the second dimension of UMAP and the volatility of Infodemic Risk Index. Panel A shows the correlation between the volatility of the Infodemic Risk Index in 2020 with a Spearman correlation of -0.30, while Panel B shows a more pronounced relationship with a Spearman correlation of -0.43 in 2021.}\label{fig6}
\end{figure}

Our final analysis examined the relationship between infodemic volatility, measured as the standard deviation of the Infodemic Risk Index, and socioeconomic factors, namely the two-dimensional socio-economic space derived from UMAP (Figure \ref{fig6}). This approach allowed us to investigate whether countries with different socioeconomic profiles exhibited distinct patterns of stability or fluctuation in their exposure to misinformation over the course of the pandemic.

We observed a pronounced negative correlation between volatility and the second dimension of the UMAP socioeconomic projection, which was consistent across both 2020 and 2021. In 2020, this correlation amounted to -0.304 with a marginally significant p-value of 0.069, while in 2021 it strengthened to -0.435 with a highly significant p-value of 0.008. This intensification of the relationship over time suggests that as the pandemic progressed, the stabilizing effect of robust institutional frameworks on information environments became more pronounced.

These results indicate that countries positioned higher along the second socioeconomic dimension, which captures aspects of institutional quality, democratic function, and media environment, exhibited significantly more stable information landscapes during the crisis. Conversely, countries lower on this dimension experienced greater fluctuations in their infodemic risk, suggesting that institutional strength may serve as a buffer against volatile shifts in information quality during prolonged crises. This finding highlights the importance of institutional frameworks not only in reducing overall infodemic risk but also in maintaining consistent, reliable information environments that can withstand external pressures and evolving crisis conditions.

\section{Discussion}\label{sec12}

Our analysis reveals complex, multidimensional relationships between socioeconomic factors and infodemic dynamics during the COVID-19 pandemic. By examining these relationships through several complementary analytical approaches, we have uncovered patterns that provide significant insights into how societal characteristics shape information environments during global crises. In this section, we interpret our findings in detail, addressing each of our five research questions while connecting our results to existing literature and drawing out theoretical implications and practical applications for policy interventions.

\subsection{Socioeconomic dimensions and country clustering}

Addressing our first research question about the underlying dimensions in socioeconomic indicators and country clustering patterns, the UMAP visualization presented in Figure 1 reveals a striking dimensional structure in our OECD data. The arrangement of countries (left panel) shows clear geographic and cultural clustering patterns, with Scandinavian and Central European nations positioned in the upper right, Anglosphere countries to the right, and several South American countries in the bottom left. This clustering reflects deeper similarities in social, political, and economic structures that transcend simple geographic proximity, supporting theories of path-dependent institutional development \cite{north1990institutions} \cite{acemoglu2005institutions} and varieties of capitalism \cite{hall2009varieties}.

More revealing is the arrangement of socioeconomic indicators (right panel) along a curved distribution. The positioning of welfare indicators and median age at the top, followed by democracy, education, and economic development indicators, with social and political polarization measures distinctly separated in the bottom left, suggests a multidimensional structure of social development. This structure aligns with Welzel and Inglehart's \cite{welzel2005democratization} theory of value change, which posits that socioeconomic development follows patterns that move from survival values to self-expression values, with corresponding shifts in institutional arrangements.

The separation of polarization indicators from other socioeconomic measures is particularly noteworthy. It suggests that polarization represents a distinct dimension of social organization that can vary independently of traditional development indicators, a finding that resonates with recent research showing how even advanced democracies can experience significant polarization under certain conditions \cite{mccoy2018polarization} \cite{boxell2024cross}. This dimensional independence helps explain why infodemic risk does not correlate straightforwardly with conventional development metrics, as nations with similar levels of economic development or democratic institutions can nevertheless experience very different levels of social and political division.

The effectiveness of UMAP in capturing these complex relationships demonstrates the value of dimensionality reduction techniques for understanding multifaceted socioeconomic data. Rather than analyzing individual indicators in isolation, this approach reveals underlying structures that better reflect the interconnected nature of social, political, and economic development.

\subsection{Temporal infodemic patterns and country clusters}

In response to our second research question regarding distinct country clusters based on temporal patterns of infodemic risk, Figure 2 presents perhaps our most intriguing finding: the emergence of two distinct clusters of countries based on their infodemic temporal patterns. The smaller cluster of nine countries (Belgium, Chile, France, Haiti, Israel, Nigeria, Singapore, Turkey, and Venezuela) displays markedly different infodemic dynamics compared to the larger group. This clustering cuts across traditional categories of economic development, geographic region, and political systems, suggesting that infodemic vulnerability follows patterns not easily reducible to conventional development metrics.

The temporal analysis in Figure \ref{fig3} further illuminates these differences, showing that while Cluster 2 countries maintained relatively stable infodemic risk levels, Cluster 1 countries experienced a dramatic spike in early 2021. This timing coincides with the global rollout of COVID-19 vaccines, suggesting that vaccine-related controversies may have triggered distinct infodemic patterns in these countries. In particular, vaccine misinformation can significantly reduce vaccination intentions, particularly in certain national contexts \cite{kim2022comprehensive} \cite{lee2022misinformation}.

The heterogeneity of countries within Cluster 1 is remarkable, including nations from different continents, political systems, and development levels. What might unite these diverse countries in their infodemic vulnerability? One possibility is that they share certain institutional features or social dynamics not captured by individual socioeconomic indicators but revealed through their information behavior patterns. For instance, they might share particular configurations of media systems, political discourse, or trust relationships that create similar vulnerabilities to infodemic spikes despite their differences in other aspects.

This finding challenges simplistic narratives about information vulnerability being primarily a function of economic development or democratic institutions. Instead, it suggests that infodemic resilience emerges from complex interactions between multiple societal factors, creating distinctive national patterns that transcend conventional categorizations. Significantly, there is a growing recognition in the literature that information disorders are shaped by context-specific institutional arrangements and historical trajectories \cite{zhu2025individual}.

\subsection{Evolving correlations between socioeconomic factors and infodemic measures}

Examining our third research question about how socioeconomic factors correlate with different measures of infodemic risk over time, Figure \ref{fig4} reveals a fascinating temporal evolution in these relationships. The first panel shows that different infodemic measures correlate with socioeconomic indicators in distinct patterns that change over time. During the pandemic's first phase (2020), the Dynamic Risk Index and Unreliability Index show stronger correlations with the first principal component of socioeconomic indicators, while in the second phase (2021 onwards), the correlation shifts to the Infodemic Risk Index.

This temporal shift suggests that different aspects of information environments become salient at different stages of a crisis. In the early, acute phase of the pandemic, socioeconomic factors appear more strongly related to dynamic measures of information engagement and sharing, perhaps reflecting how societies initially respond to novel threats with varying levels of information seeking and sharing behaviors. As the pandemic became a chronic condition, socioeconomic factors increasingly correlated with overall infodemic risk, suggesting that deeper structural characteristics of societies became more determinative of information quality as the crisis progressed.

The second panel of Figure \ref{fig4}, showing the statistical significance of these correlations, reinforces this interpretation. The p-values for Dynamic Risk Index and Unreliability Index are consistently significant in the pandemic's first year, while the Infodemic Risk Index becomes significant in the second year. This provides some statistical support for the changing relationship between socioeconomic factors and different aspects of information environments over time.

This temporal dimension has been overlooked in previous research on misinformation dynamics, which has tended to analyze phenomena at single time points rather than as evolving processes. Our findings suggest that understanding infodemics requires an attention to these temporal patterns, in particular about how information environments evolve through different phases of a crisis, and how their relationship with underlying societal factors changes over time.

\subsection{The protective role of media diet diversity}

Turning to our fourth research question regarding the role of news media diet diversity in mediating the relationship between socioeconomic factors and infodemic vulnerability, Figure \ref{fig5} presents evidence for the protective effect of information pluralism. The positive correlations between the Shannon entropy of shared web domains and both dimensions of the UMAP socioeconomic space (panels A and B) indicate that countries with more diverse information ecosystems tend to have stronger democratic institutions, higher education levels, and greater economic development. This relationship is even stronger when focusing specifically on political news (panels C and D), suggesting that exposure to diverse political viewpoints may be particularly important for information resilience.

These findings support and extend previous work suggesting that media diversity contributes to democratic health and information quality. At the individual level, higher news literacy tends to be associated to richer media diets \cite{van2021fighting} and improves the identification of fake news \cite{jones2021does}. At the societal level, media legitimacy and trust emerge as key factors in national resilience against misinformation \cite{rodriguez2023exploring}.

Our results therefore provide some support to the ``information pluralism'' hypothesis, namely the idea that more diverse information ecosystems create conditions where misinformation is less likely to dominate the public discourse. When citizens are exposed to a variety of sources and perspectives, they may develop greater critical evaluation skills and become less susceptible to misleading narratives from any single source. This challenges ``echo chamber'' theories that suggest that greater information choice necessarily leads to selective exposure and polarization \cite{pariser2011filter}. Instead, our findings are more consistent with research showing that most citizens' media diets are actually more diverse than commonly assumed \cite{guess2021almost}, and that this diversity can serve protective functions and favors moderation.

The stronger correlation observed for political news diversity is particularly interesting. It suggests that exposure to varied political viewpoints, rather than a narrow ideological information diet, may be especially important for building societal resilience to misinformation. This emphasizes the value of cross-cutting exposure in democratic discourse \cite{mutz2006hearing} and contradicts concerns that political diversity necessarily exacerbates polarization. Rather, it is polarization that poses a threat to democracy by reducing interest and information diversity \cite{bednar2021polarization}.

\subsection{Infodemic volatility and institutional stability}

Addressing our fifth research question on how infodemic volatility relates to socioeconomic factors and whether this relationship evolves over time, Figure \ref{fig6} reveals an important negative correlation between infodemic volatility (standard deviation of the Infodemic Risk Index) and the second dimension of our UMAP socioeconomic space. Countries positioned higher on this dimension, which captures aspects of institutional strength and media environment quality, demonstrate more stable information landscapes even amid the uncertainty of a global pandemic. This stability became more pronounced in 2021 (panel B, R = -0.435) compared to 2020 (panel A, R = -0.304), suggesting that institutional frameworks tend to become increasingly important for maintaining information integrity as crises evolve.

Institutional theories of resilience that emphasize how formal structures provide stability during disruptions \cite{aldrich2012building}. In the context of information disorders, robust institutions appear to help societies maintain more consistent information environments, reducing the dramatic fluctuations in reliability that characterized high-risk countries. The strengthening of this relationship over time further suggests that institutional factors become more determinative of information quality as crises progress from acute emergency to chronic challenge.

This finding also contributes to debates about what constitutes information system resilience during crises. Rather than merely looking at average levels of misinformation, our analysis suggests that volatility, i.e. the tendency of information environments to fluctuate dramatically, represents another important dimension of infodemic vulnerability. Societies with greater institutional stability appear better able to maintain consistent information environments even amid uncertainty, avoiding the dramatic swings between reliable and unreliable information that can disorient public understanding and undermine trust.

\subsection{Theoretical and policy implications}

Our findings contribute to an emerging theoretical framework that views infodemics as a complex socio-technical phenomenon, deeply embedded in broader social structures. Rather than treating misinformation as primarily a problem of individual psychology or platform design, this perspective emphasizes how historical institutional developments, social trust, educational systems, and political cultures create conditions that either foster or inhibit information disorders during crises.

The multidimensional nature of socioeconomic factors revealed in our UMAP analysis suggests that simple models of information vulnerability based on single indicators (such as education levels or economic development) are insufficient. Instead, countries occupy positions in a complex socioeconomic space with multiple dimensions that interact to create distinctive patterns of vulnerability or resilience. The emergence of distinct clusters based on temporal infodemic patterns, cutting across traditional development categories, suggests that information vulnerability follows its own logic, not entirely independent from other socioeconomic factors, but not reducible to them either. This challenges both technological determinism (the idea that digital media platforms inherently produce certain information outcomes regardless of context) and simple socioeconomic determinism (the notion that information quality mechanically reflects development levels).

The changing correlations between socioeconomic factors and different infodemic measures over time point toward a dynamic model of information disorders, where relationships between structural conditions and information environments evolve through different phases of a crisis. This temporal dimension has been undertheorized in most existing frameworks, which tend to present static models of misinformation vulnerability.

Finally, the strong relationship between news media diet diversity and positions in socioeconomic space suggests a possible causal pathway, that needs to be further investigated in future research, through which broader societal characteristics influence information environments. Societies with certain institutional arrangements, educational systems, and political cultures may foster more diverse information ecosystems, which in turn provide greater resilience against infodemics during crises.

Our findings have significant implications for policy approaches to managing infodemics and building societal resilience against information disorders. We can identify several key areas where policy interventions might be most effective, based on our empirical results.

As to media diversity and pluralism, the strong correlation between news media diet diversity and reduced infodemic risk suggests that policies promoting media pluralism could enhance societal resilience against misinformation. Policy approaches should support diverse media ecosystems through funding for public service media, subsidies for local news, and anti-monopoly regulations in media markets. Additionally, developing media literacy programs that encourage citizens to consume information from multiple sources would strengthen individual information evaluation skills. Creating incentives for cross-cutting exposure to diverse political viewpoints and implementing platform design features that promote exposure to diverse content could further enhance information ecosystem health. The particularly strong relationship between political news diversity and socioeconomic indicators associated with lower infodemic risk suggests that ensuring pluralism in political information sources may be especially important. This challenges approaches focused primarily on removing ``harmful'' content, suggesting instead that ensuring a healthy diversity of perspectives may be more effective for long-term resilience.

Concerning the strengthening of the institutional framework, the negative correlation between infodemic volatility and the institutional dimension of our socioeconomic space indicates that stronger institutions help stabilize information environments during crises. This suggests policies aimed at building trust in key epistemic institutions such as science, journalism, and public health authorities. Supporting independent fact-checking organizations and quality journalism provides additional safeguards against the spreading of misinformation. Designing institutional arrangements that aim at reducing political polarization can mitigate one of the key drivers of information disorder. Furthermore, creating rapid response systems that can provide authoritative information during emerging crises enables a more timely correction of misinformation before it gains traction. The strengthening of this relationship over time suggests that institutional factors become increasingly important as crises progress from acute emergency to chronic challenge. Institutional development inevitably paints with a long-term investment in information resilience rather than with the search for quick fixes to immediate information disorders.

The distinct separation of polarization indicators from other socioeconomic measures in our UMAP analysis suggests that this dimension requires specific policy attention. Countries may have well-developed economies, strong democratic institutions, and high education levels, yet still experience high levels of polarization that create vulnerability to infodemics. This makes a case for policies focused on reducing affective polarization through interventions that humanize political outgroups. Additionally, developing political institutions that incentivize cooperation rather than zero-sum competition could mitigate some of the structural drivers of polarization. Creating shared information spaces that transcend partisan divides would provide common ground for diverse groups to engage with the same factual basis. Supporting trust-building initiatives across political and social boundaries further addresses the social-psychological dimensions of polarization. The distinct clustering of polarization indicators separate from traditional development metrics suggests that these interventions may be necessary even in highly developed societies with otherwise strong institutions.

The changing correlations between socioeconomic factors and different infodemic measures over time suggest that effective interventions may need to adapt to different phases of a crisis. During early, acute phases, when dynamic information sharing behaviors are more strongly correlated with socioeconomic factors, interventions should focus on the rapid dissemination of accurate information through trusted channels. Supporting accurate information sharing behaviors among influential nodes in social networks becomes crucial in these initial stages. Providing clear guidance to reduce uncertainty can prevent information vacuums that misinformation often fills. As crises become chronic, and overall infodemic risk becomes more strongly correlated with socioeconomic factors, interventions might shift toward addressing deeper structural factors that shape information environments. Building long-term resilience through educational initiatives enables preparation for future information challenges. This temporal dimension suggests that infodemic management requires adaptive response strategies rather than one-size-fits-all approaches.

\subsection{Limitations and future directions}

Our results are subject to several limitations. First of all, our reliance on Twitter data implies that our analysis captures only a portion of the total information ecosystem, with known demographic biases toward younger, more educated, and more politically engaged populations \cite{mellon2017twitter}. Considering more data sources, including other social media platforms, traditional media, and offline communication channels, is essential to provide a more comprehensive view of information flows across different demographic segments.

Moreover, the cross-sectional nature of our socioeconomic indicators limits causal inferences about their relationship with infodemic patterns. While we observe correlations between socioeconomic factors and infodemic measures, we cannot establish causal pathways, and our preliminary results should be seen mainly as a useful source of testable predictions for future longitudinal studies tracking changes in both socioeconomic factors and information environment characteristics over time.

Additionally, our focus on country-level analysis necessarily obscures within-country variations that may be substantial, particularly in larger or more diverse nations. Regional disparities in education, economic development, political attitudes, and media access likely create uneven information landscapes within countries. A closer focus on how regional differences in socioeconomic factors within countries relate to more localized patterns of infodemic spread is therefore needed.

Our analysis includes a comprehensive but not exhaustive set of socioeconomic indicators. Some potentially relevant factors, such as historical experiences with epidemics, cultural attitudes toward authority, or specific features of healthcare systems, were not included in our dataset. Enriching the panel of socioeconomic idicators with additional relevant dimensions that might influence infodemic vulnerability would be another major step forward.

Finally, while our study identifies correlations between socioeconomic factors and infodemic patterns, an understanding of the precise mechanisms through which these factors influence information environments is still largely uncharted territory. Mixed-methods approaches combining quantitative analyses with qualitative case studies could be useful for hypothesis development in this respect. Comparative case studies of countries with similar socioeconomic profiles but different infodemic outcomes would also be particularly valuable for identifying specific institutional arrangements or cultural factors that enhance infodemic resilience.

\section{Conclusion}

Our study demonstrates that the vulnerability of societies to infodemics during the COVID-19 pandemic was significantly shaped by underlying socioeconomic factors, with patterns that evolved over time as the crisis progressed. The arrangement of socioeconomic indicators in a multidimensional space, the clustering of countries based on temporal infodemic patterns, the changing correlations between socioeconomic factors and infodemic measures over time, and the relationship between news media diet diversity and infodemic risk all point toward a complex, context-dependent model of information vulnerability during crises.

These findings challenge simplistic narratives about information disorders seen as primarily technological problems related to digital platform design and functioning, or straightforward reflections of the levels of economic development. Instead, they suggest that infodemics emerge from complex interactions between technological systems, social structures, institutional arrangements, and cultural factors, creating distinctive patterns of vulnerability and resilience across different societies.

The diversity of media diets, particularly regarding political information, emerged as an especially important protective factor against information disorder, highlighting the value of pluralistic information ecosystems for societal resilience. The negative correlation between institutional stability and infodemic volatility further emphasizes the importance of robust institutions for maintaining consistent, reliable information environments during crises.

These insights underscore the need for more comprehensive approaches to infodemic management that address not only immediate manifestations of misinformation but also the deeper structural factors that create favorable conditions for information disorders. Building societal resilience against future infodemics will require coordinated efforts across educational, institutional, and media domains, guided by a nuanced understanding of how socioeconomic factors shape information environments during crises.

\section*{Acknowledgements}

We acknowledge the contribution of Manlio De Domenico, who led the COVID-19 Infodemic Observatory project that enabled the collection of the dataset analyzed in this work. R.G. acknowledges the financial support received from the European Union’s Horizon Europe research and innovation program under grant agreement No 101070190. 

\section*{Declarations}

%Some journals require declarations to be submitted in a standardised format. Please check the Instructions for Authors of the journal to which you are submitting to see if you need to complete this section. If yes, your manuscript must contain the following sections under the heading `Declarations':

\begin{itemize}
%\item Funding
%\item Conflict of interest/Competing interests (check journal-specific guidelines for which heading to use)
%\item Ethics approval and consent to participate
%\item Consent for publication
\item Data availability 
Data analyzed in this work are available from the corresponding author upon reasonable request.
\end{itemize}

\begin{appendices}

\section{Appendix A}\label{secA1}

\renewcommand{\arraystretch}{1.2} % Aumenta lo spazio tra le righe
\setlength{\tabcolsep}{5pt} % Imposta la spaziatura tra colonne
\rowcolors{2}{gray!15}{white} % Colora le righe alternate

\begin{longtable}{p{4cm} p{11cm}}
    \caption{List of the socioeconomic indicators considered for analysis Table of the socioeconomic indices, the organization and sources, and a brief description.} \\
    \toprule
    \rowcolor{gray!40} \textbf{Variable} & \textbf{Description} \\
    \midrule
    \endfirsthead

    % Intestazione per le pagine successive
    \toprule
    \rowcolor{gray!40} \textbf{Variable} & \textbf{Description} \\
    \midrule
    \endhead

    % Righe della tabella
    eiu democracy index & It measures the level of democracy in a given country, based on 60 different indicators over 5 main categories: electoral processes and pluralism, functioning of government, political participation, democratic political culture, and civil liberties. It is produced by the Economist Intelligent Unit (EIU) of the Economist group. The data come from experts’ assessments and public opinion surveys, mainly from the World Values Survey. The index ranges from 0 (highly undemocratic) to 10 (highly democratic) \cite{eiuDemocracyIndex}. \\
    eiu political participation & The EIU Political Participation indicator \cite{un2020pop} measures the degree of citizen engagement in political processes. It corresponds to one of the 5 main categories of indicators used to compute the Economist Democracy Index. It examines factors such as voter turnout, citizens’ engagement with politics, party membership and participation in civil society organizations.  \\
    fh freedom index & The Freedom Index \cite{freedomhousePandemicDigital}, produced by the US NGO Freedom House, assesses the political freedoms and civil liberties of citizens in a given country. It ranges from 0 to 100 and is composed of 10 political rights indicators and 15 civil liberties indicators which assess several dimensions of freedom, including electoral processes, political pluralism, freedom of expression and associational rights. \\
    iep global peace index & The Global Peace Index \cite{gdi2021global}, compiled by the Institute for Economics and Peace (IEP), quantifies the relative peacefulness of countries and regions. It incorporates various indicators, including levels of violence, militarization, and political instability, to assess the overall peacefulness and stability of a nation. Originally, the index was expressed on a scale from 1 to 5, where 1 signifies most peaceful and 5 least peaceful. In order to align its interpretation to that of the other indices, it was reversed. \\
    iep global terrorism index &  The Global Terrorism Index \cite{vision2020gti}, also produced by the Institute for Economics and Peace (IEP), provides a comprehensive measure of terrorism-related activities worldwide. It analyzes factors such as the number of terrorist incidents, casualties, and the economic impact of terrorism, enabling a comparative assessment of terrorism levels across countries. \\
    imf gdp percap current prices & This variable corresponds to the Gross Domestic Product per capita at current prices, as reported by the International Monetary Fund \cite{imfWorldEconomic}. It is calculated by dividing the GDP by the population size and serves as a measure of the economic prosperity and standard of living within a nation. \\
    oecd perc adults tertiary edu & This variable represents the percentage of adults (aged 25-64) with tertiary education in a given OECD country \cite{oecd2023Tertiary}. \\
    oecd perc foreign students & This variable measures the percentage of foreign students enrolled in educational institutions in a given OECD country. It offers insights into a nation’s attractiveness as an educational destination and reflects its openness to international students.\\
    oecd pisa mathematics & This variable represents the performance of 15-year-old students in mathematics, as assessed by the 2018 Programme for International Student Assessment (PISA). Provides a measure of educational results in each of the 78 countries surveyed \cite{oecd2023pisa}. \\
    oecd pisa science & This variable represents the scientific knowledge and skills of 15-year-old students in a given country, as assessed by the 2018 Programme for International Student Assessment (PISA) \cite{oecd2023pisa}. \\
    oecd soc security & This indicator measures the extent of social security contributions within a given OECD country as a percentage of its GDP. These are “compulsory payments paid to general government that confer entitlement to receive a (contingent) future social benefit” and provide insights into the level of social protection within a country, its economic development and the level of social inequality \cite{oecd2023socsec}. \\
    oecd social spending & This variable quantifies the share of a given OECD country’s GDP allocated to social expenditures, including welfare, healthcare, and family support programs. It serves as an indicator of the country’s level of social development and its commitment to redistributing resources for the well-being of its population \cite{oecd2023socspending}. \\
    oecd trust government & This variable assesses the share of citizens of a given country that report to trust their own country’s government. It is calculated using the number of people answering positively to the survey question “In this country, do you have confidence in the national government?” \cite{oecd2023trust_gov}. \\
    rsf freedom press & The Press Freedom Index evaluates the level of freedom enjoyed by journalists and the media in a given country. It is produced by the France-based NGO Reporters Without Borders. It assesses factors such as media independence, access to information, and protection of journalists, providing insights into the state of press freedom and freedom of expression \cite{rsf2020World}. \\
    un egov e-particpation index & The E-Participation Index measures the level of citizen participation in online government processes and decision-making. It is calculated in the United Nations E-Government Survey carried out by the UN Department of Economic and Social Affairs. It assesses the effectiveness of IT services that facilitate information exchange and interaction between government and citizens. Moreover, it gives a measure of citizen involvement in policy and decision-making. The index ranges from 0 to 1 \cite{un2020egov}. \\
    un median age & This variable measures the median age of a country, namely the age that divides a population into two equal halves, with half of the population being older and half being younger than that. The estimates are provided by the United Nations \cite{un2020pop}. \\
    vdem pol polarization & This indicator measures “the extent to which political differences affect social relationships beyond political discussions” \cite{vdem2022varieties}. When supporters of contrasting political factions are hesitant to participate in friendly interactions, which could happen in a variety of contexts such as in their workplace or in family functions, it indicates a high level of political polarization within societies. This variable is derived from the V-Dem (Varieties of Democracy) dataset, a large-scale political science dataset which aggregates country expert judgments in order to produce valid estimates of several concepts related to democracy. In particular, this indicator ranges from 0 to 4, where 0 signifies no polarization (i.e., supporters of opposing political factions tend to engage with each other in a cordial and friendly manner) and 4 represents the highest amount of polarization, with a
generally hostile social environment among supporters of different political factions. \\
vdem soc polarization & This variable measures the level of polarization in a society. Unlike the previous variable, which focused on the nature of interactions between supporters of opposing political camps, this indicator provides information on whether 'there is general agreement on the general direction this society should develop' \cite{vdem2022varieties}. More specifically, this variable comes from the aggregation of country experts answering the question “How would you characterize the differences of opinions on major political issues in this society?”. The possible values originally span from 0 to 4, with 0 signifying serious polarization in society, resulting in major clashes of views, and 4 meaning that, despite the presence of differing opinions, there is a general consensus on the direction for key political issues. However, similarly to the Global Peace Index, the values of this index were flipped to achieve easier interpretation. Consequently, higher values of the metric in the final dataset
correspond to higher levels of polarization. \\
wb perc internet use & This indicator measures the level of Internet use in a given
country, according to World Bank estimates \cite{worldbank2020internet}. In particular, it represents the percentage of individuals who reported in a survey to have used the Internet in the last 3 months. It provides insights into the level of internet penetration, digital connectivity and technological development in a country.\\
wb gini coefficient & The Gini coefficient is a measure of income inequality. It ranges from 0 to 1, where higher values represent greater income inequality. Data come from the World Bank, except in the case of New Zealand, where the estimate is produced by the OECD \cite{worldbank2020gini}. \\
    
    \bottomrule
\end{longtable}

\end{appendices}

\bibliographystyle{plain}
\bibliography{references}

\end{document}